# Analog CMOS-based Resistive Processing Unit for Deep Neural Network Training

Seyoung Kim[*], Tayfun Gokmen[*], Hyung-Min Lee[†] and Wilfried E. Haensch[*]
[*]IBM Thomas J. Watson Research Center, Yorktown Heights, New York, USA
[†]School of Electrical Engineering, Korea University, Seoul, South Korea
E-mail: sykim@us.ibm.com

*Abstract*—**Recently we have shown that an architecture based on resistive processing unit (RPU) devices has potential to achieve significant acceleration in deep neural network (DNN) training compared to today's software-based DNN implementations running on CPU/GPU. However, currently available device candidates based on non-volatile memory technologies do not satisfy all the requirements to realize the RPU concept. Here, we propose an analog CMOS-based RPU design (CMOS RPU) which can store and process data locally and can be operated in a massively parallel manner. We analyze various properties of the CMOS RPU to evaluate the functionality and feasibility for acceleration of DNN training.**

## I. INTRODUCTION

Deep neural network (DNN) [1] based models have made significant progress in the past few years due to the availability of large labeled datasets and continuous improvements in computation resources. Some of these exceed human level performance on various tasks such as object/speech recognition, language translation and image captioning. Since the quality of the models depends on large amount of training data and increased complexity of the neural network, training remains a bottleneck. In some cases it may take weeks of time even on parallel and distributed computing frameworks utilizing many computing nodes [2]–[4] to complete a model. To reduce training time, hardware acceleration for these workloads has been pursued either in conventional CMOS technologies or by using emerging non-volatile memory (NVM) technologies [5].

DNN training generally relies on the backpropagation algorithm which is composed of three repeating cycles: forward, backward and weight update [6]. In our recent publications [7], [8], we have proposed an architecture based on 2D crossbar array of resistive processing unit (RPU) devices which can perform all three cycles of the backpropagation algorithm in parallel, thus potentially providing significant acceleration in DNN training with lower power and reduced computation resources compared to today's CPU/GPU implementations

To build such hardware and achieve large acceleration, it is critical to find the suitable RPU element. Appropriate device specifications for these RPU elements were found by a systematic analysis of the technology-relevant parameters and their influence on the training process. The results of this evaluation are shown in Table I [7] and can be summarized as follows: A RPU cell is a resistor-like circuit component with tunable conductance which, for example, can switch its conductance state by 1-ns pulses. The dynamic range should cover at least 1,000 steps to switch from the lowest conductance state to the highest in an analog and incremental manner. One order of magnitude conductance contrast between max and min conductance state (on/off ratio) is sufficient, and the device resistance of 24 MΩ is required for a 4,096 × 4,096 device array. The conductance increase and decrease (up and down) need to be symmetric within 5% mismatch error.

TABLE I.  DESIRED RPU CELL PROPERTIES

| Specifications | Parameter | Value |
|---|---|---|
| Storage capacity | $(\max(g_{ij}) - \min(g_{ij}))/\Delta g_{min}$ | 1000 levels |
| Update pulse duration | $t_{pulse}$ | 1 ns |
| Average device resistance | $R_{device}$ | 24 MΩ |
| On/off ratio | $\max(g_{ij})/\min(g_{ij})$ | 8 |
| Up/down symmetry | $\Delta g_{min}^+/\Delta g_{min}^-$ | 1.05 |
| Device area | $W \times L$ | 0.04 μm² |

Conventional non-volatile memory elements do not fulfill all of these requirements, and material research is needed to explore the RPU concept. Here, we propose an analog CMOS-based RPU (CMOS RPU) cell which is designed with conventional circuit components to simulate the desired RPU device specifications. We perform circuit and DNN training simulations to demonstrate the functionality and test feasibility of the CMOS RPU for DNN training acceleration. We also discuss the non-ideal characteristics of the CMOS RPU, and compare it with ideal RPU specifications.

## II. ANALOG CMOS RPU DESIGN

### A. RPU System

An RPU system utilizing a crossbar array of RPU cells for DNN computation is shown in Fig. 1. The conductance of each RPU cell represents a matrix element or weight, which can be updated or accessed through peripheral circuits. A block of identical peripheral circuits is connected to each row and column in the array to perform forward, backward and update operations of the backpropagation algorithm. The key functions of the peripheral circuit block are shown in Fig 3 and 4: (i) to apply pulse-width modulated (PWM) read pulses depending on the input vector during forward/backward cycles, (ii) to integrate read current accumulated from the connected cells and convert the integrated current into a digital value for subsequent computation, and (iii) to apply a series of stochastically populated pulses based on the supplied update vectors for weight update cycles and.

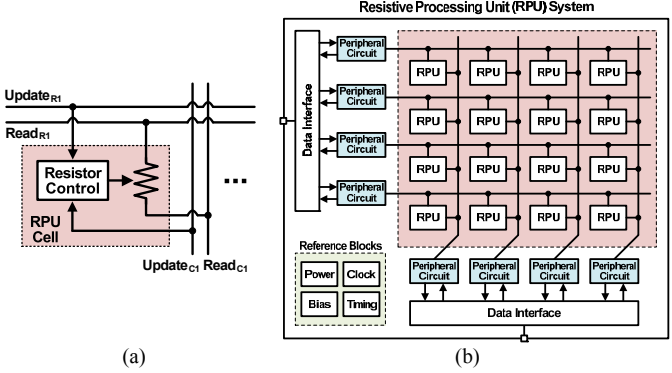

Fig. 1. (a) Conceptual diagram of the RPU cell. The cell conductance, i.e. a weight value, can be adjusted and sensed through update and read lines, respectively. (b) Block diagram of the resistive processing unit (RPU) system.

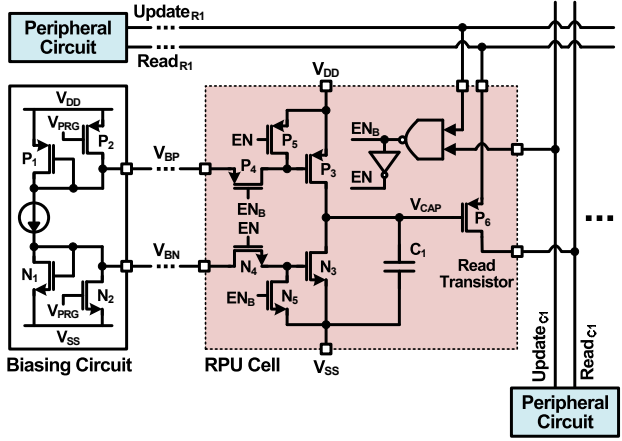

Fig. 2. Schematics of the analog CMOS RPU cell and shared biasing circuit.

### B. CMOS RPU cell

Based on the specifications and operation principles for the RPU device, we implement an analog CMOS-based RPU with conventional circuit components, a capacitor and a set of transistors, as shown in Fig. 2. In this design, the capacitor, $C_1$, serves as a memory element in the cell and stores the weight value in the form of electric charge. The capacitor voltage, $V_{cap}$, is directly applied to the gate terminal of the *read transistor*, P6, and modulates its channel resistance, $R_{on,P6}$. Therefore, the charge state stored in the capacitor can be accessed by applying small bias across P6 and measuring the current, $I_{read}$. For forward and backward (read) operation, voltage pulses with predefined amplitude, e.g. $V_{DD}$ and $V_{DD}$ - 0.1, are applied to P6 from $Read_{R1}$ and $Read_{C1}$ lines. The small source-drain voltage leads P6 to operate in a triode region which on-resistance depends on the gate voltage of P6 as,

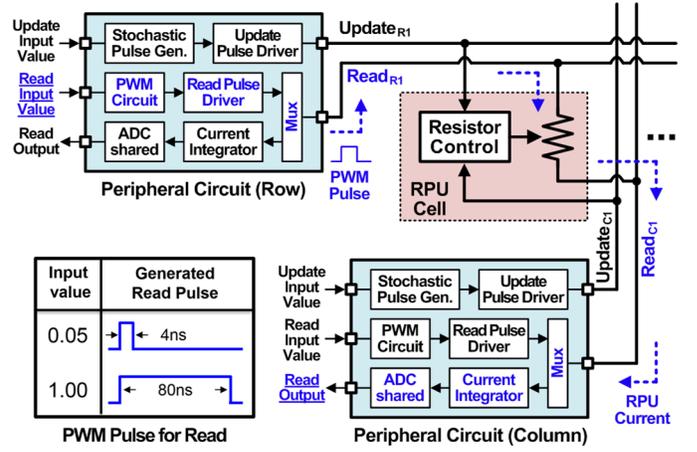

Fig. 3. Block diagram of the RPU cell and peripheral circuits with emphasis on the read operation used for forward and backward cycles.

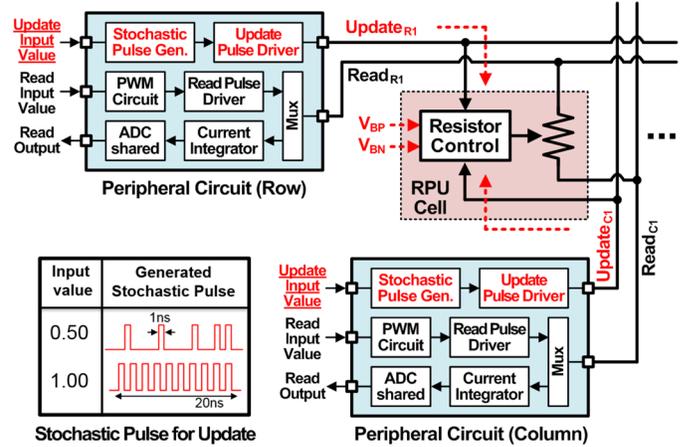

Fig. 4. Block diagram of the RPU cell and peripheral circuits with emphasis on the update operation to control the RPU resistance.

$$R_{on,P6} = \frac{V_{SD,P6}}{I_{D,P6}} = \frac{1}{\mu_p C_{ox} \frac{W}{L}(V_{DD} - V_{CAP} - V_{TH,P6})} \quad (1)$$

The read currents from read transistors in RPU cells are accumulated through either the $Read_{C1}$ line (forward) or $Read_{R1}$ line (backward) and integrated in the peripheral circuits to accomplish the vector-matrix multiplication.

For the weight update, a stochastic computing scheme [7] is used where the local multiplication operation is performed by using a simple coincidence detection method. With the NAND and inverter logics connected with $Update_{R1}$ and $Update_{C1}$, the stored weight in a CMOS RPU cell can be updated only when two stochastic pulses from update lines (row and column) are coinciding. For example, when both $Update_{R1}$ and $Update_{C1}$ signals are high for update operation (i.e. EN is high), current source transistors, P3 and N3, receive bias voltages, $V_{BP}$ and $V_{BN}$, from a global biasing circuit through switches, P4 and N4, respectively. Then, either P3 or N3 provides charging or discharging current to $C_1$, respectively, depending on a programming polarity signal, $V_{PRG}$, in a biasing circuit, which controls the $V_{cap}$ level and thus resistance of P6.

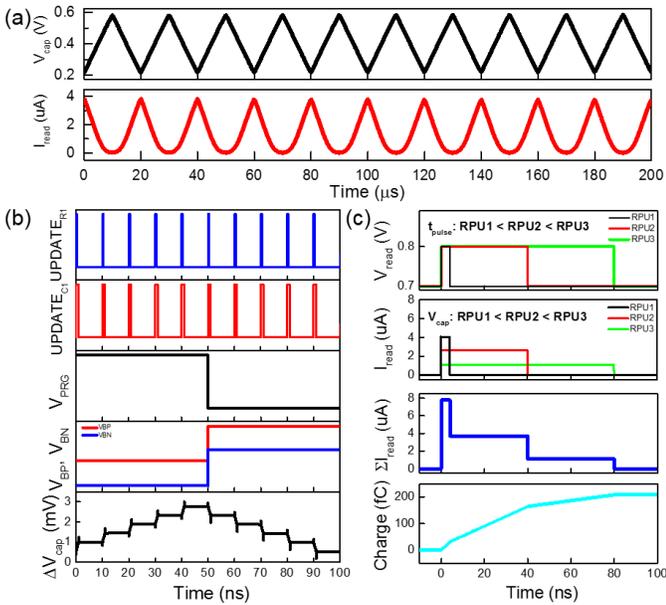

Fig. 5. Update and read operation in the CMOS RPU cell. (a) $V_{cap}$ and $I_{read}$ vs time when 1,000 up and 1,000 down update pulses with a 1-ns pulse width are given to the update lines for 10 cycles. Simulated waveform for (b) update operation to control the conductance of the RPU cell and (c) read operation using a 3 × 1 RPU array.

## III. RESULTS AND DISCUSSION

To confirm the functionality of CMOS RPU cell, we perform SPICE simulation and prove that the proposed CMOS RPU cell design in Fig. 2 meets most of the specifications in Table I except the area and device resistance.

### A. Analog and incremental weight update

Linear and incremental $V_{cap}$ update and corresponding $I_{read}$ change are displayed as a function of time in Fig. 5a. Alternating sets of 1,000 up pulses and down pulses with 1-ns pulse width are applied for 10 cycles. For weight update, waveforms of update lines, $V_{PRG}$, $V_{BP}$, $V_{BN}$, and consequent voltage change at $C_1$ (= $\Delta V_{cap}$) are displayed for five charging and discharging cycles with a 10-ns period per cycle in Fig. 5b. A global $V_{PRG}$ signal determines the charging (when $V_{PRG} = V_{DD}$) and discharging (when $V_{PRG} = V_{SS}$) cycle. We expect the weight changes only when pulses at UPDATE$_{R1}$ and UPDATE$_{C1}$ lines coincide to each other. 1-ns pulse for UPDATE$_{R1}$ and 2-ns pulse for UPDATE$_{C1}$ are used in this simulation for visualization. DC bias values at $V_{BP}$ and $V_{BN}$ are determined to control and match the charging and discharging currents. The following equation determines the amount of the minimum voltage change per update:

$$\Delta V_{cap,min} = \frac{I_{D,P3,N3} \cdot t_{pulse,min}}{C_1} \quad (2)$$

where $I_{D,P3,N3}$ is the charging and discharging currents from P$_3$ and N$_3$, respectively, and $t_{pulse,min}$ is the minimum programming pulse width. The uniform step for each 10-ns update period (5 up and 5 down steps) is observed both for charging and discharging cases, reproducing the expected update behavior of an RPU device. We note that the observed non-linear dependence of the $V_{cap}$ on $I_{read}$ in Fig. 6a corresponds to a non-linear weight change shown in [7] that it does not affect training accuracy as long as up and down updates are symmetric.

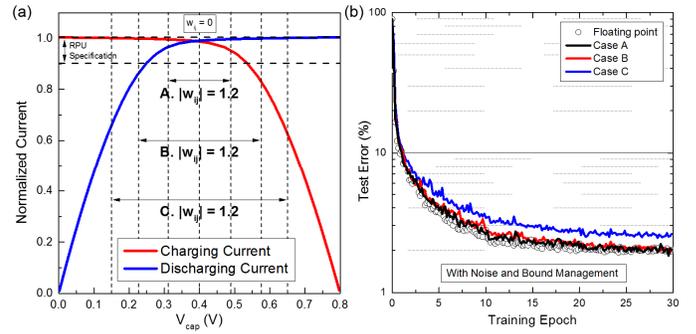

Fig. 6. Up/down update symmetry. (a) Simulated charging/discharging current vs $V_{cap}$ showing the origin of asymmetry in up and down weight update and (b) impact of the asymmetry on DNN model error rate.

### B. Read operation and device resistance

Read operation for forward and backward cycles is demonstrated using a 3 × 1 CMOS RPU array, and the waveform is summarized in Fig. 5c. Read voltage pulses with 4-ns, 40-ns and 80-ns duration, as an example, are applied to each read line in the row, and $I_{read}$ values are monitored at the column read lines. The read currents with different magnitude and duration are accumulated by the shared read line in the column and finally integrated to complete the vector-matrix multiplication. We note that $I_{read}$ range measured at $V_{SD,P6} = 0.1$ V shown in Fig. 5a and 5c indicates lower device resistance than 24 MΩ defined in RPU specification. The low read resistance issue can be addressed by using a long-channel and low-mobility FET.

### C. Up/down update symmetry

One of the most challenging specifications for current RPU device implementations is the up/down update symmetry. In CMOS RPU, the up/down symmetry can be achieved by tuning $V_{BP}$ and $V_{BN}$ to match the charging and discharging currents, $I_{D,P3}$ and $I_{D,N3}$, respectively. As shown by SPICE simulations in Fig. 6a, charging/discharging current matches well near the center voltage, 0.4 V, and starts to deviate as $V_{cap}$ approaches to the $V_{DD}$ or $V_{SS}$ due to the operation mode change in the current source transistors. To test the impact of the observed asymmetry, we performed DNN training as shown in Fig 6b for the same network described in [7] with the noise and the bound management techniques described in [8]. Cases A, B and C correspond to mapping of the identical weight range, |w$_{ij}$| = 1.2, to different voltage ranges, 0.16V, 0.24V and 0.48V, respectively. Case C uses the largest voltage range to map the weight value and, therefore, has the most severe asymmetry throughout the weight range. This model clearly deviates from the baseline model using floating point numbers. In contrast, models using a smaller voltage range, as illustrated in Cases A and B, show low test error rates comparable to the base line model. Although mapping an arbitrarily small voltage range makes the updates more symmetric, it eventually starts to affect the read accuracy on the read transistor because of the limited voltage range. Because of these tradeoffs it is ideal to use the largest acceptable voltage range as illustrated by Case B.

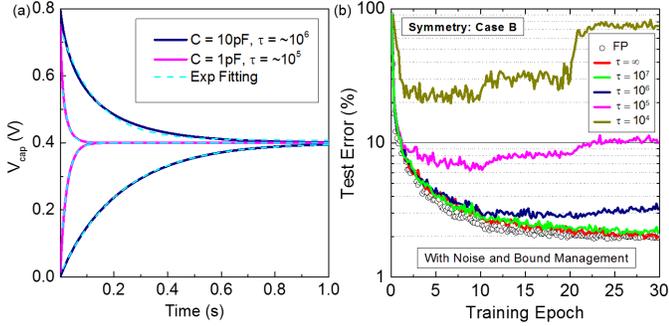

Fig. 7. Weight retention characteristics in the CMOS RPU cell. (a) $V_{cap}$ decay as a function of time due to the charge loss in the capacitor by various leakage current paths in the CMOS RPU cell, and (b) its impact on DNN model error rate.

*D. Retention*

In contrast to the NVM-based RPU cells, the CMOS RPU cell has an intrinsic retention issue due to the charge loss in the capacitor by leakage currents. In order to test the impact of retention on performance, we performed DNN training using Case B condition in Fig. 6b as a basis and introduced exponential decay term on the weights, with a time constant, $\tau$. The usage of exponential decay term is justified by the observed decay behavior in SPICE simulation results in Fig. 7a. The DNN training results are summarized in Fig. 7b where $\tau$ values are reported in the units of the time to train a single image (total duration for one set of forward, backward and update cycles). The model with $\tau = 10^4$ shows a higher test error rate although the network progresses to a level of ~ 20% error rate for the first 10 epochs when learning rate is 0.01. As the learning rate is reduced by a factor of 2 at $10^{th}$ and $20^{th}$ epochs, however, weight decay process becomes more dominant, and the error rate starts to increase. Even for a model with $\tau = 10^6$, this reduction in the learning rate creates a visible fluctuation, and the error rate climbs beyond $20^{th}$ epoch. It is clear that the weight update process is fighting against the decay process, and for competitive error rates, $\tau$ has to be at least $10^6$. Assuming a 200-ns training time per image [7], the average retention time requirement is 0.2 seconds. We emphasize that this is an average value and the training accuracy is insensitive to the variations on this parameter up to a value of 30% (data not shown).

We note that a choice of the low leakage technology and advanced biasing techniques can relieve the negative impact of relatively short retention time. The dominant leakage source is identified to be the off-state currents through the channels of $P_3$ and $N_3$ in Fig. 2. For forward and backward cycles, EN becomes low, and switches, $P_5$ and $N_5$, turn off the current sources, $P_3$ and $N_3$, respectively, to hold the $V_{cap}$ level in $C_1$. For longer retention period, leakage currents affecting $V_{cap}$ can be further reduced by applying positive and negative gate-source voltages to further turn off the current source transistors, $P_3$ and $N_3$, respectively.

TABLE II. CMOS RPU CELL SPECIFICATION

| Specifications | Ideal value | CMOS RPU |
|---|---|---|
| Storage capacity | > 1000 levels | Achieved |
| Update pulse duration | 1 ns | Achieved |
| Average device resistance | 24 MΩ | **Achievable** |
| On/off ratio | 8 | Achieved |
| Up/down symmetry | 1.05 | Achieved |
| Device area | 0.04 μm$^2$ | **Larger** |

*E. Size and area*

The capacitor, $C_1$, needs to be large enough to guarantee sufficient retention time, but it typically dominates the area of the CMOS RPU cell. To minimize the capacitor area in the RPU cell, high-density deep trench capacitors for embedded DRAM technology were used. Further design optimization and modification may be required to minimize the cell area while sufficing other requirements.

IV. CONCLUSION

We proposed an analog CMOS-based RPU cell design that fulfills most of the challenging requirements for RPU devices as summarized in Table. II, which have not been achieved by other implementations to date. We discussed the operation principles of the CMOS RPU in the crossbar array configuration and demonstrated key functionality for DNN training applications. CMOS RPU-specific parameters, which can impact the accelerator performance, are identified and the impact is addressed by performing DNN training simulations using hardware defined constraints.

ACKNOWLEDGMENT

We thank Yulong Li and Paul Solomon for useful discussions and suggestions.

REFERENCES

[1] Y. LeCun, Y. Bengio, and G. Hinton, "Deep learning," *Nature*, vol. 521, no. 7553, pp. 436–444, May 2015.
[2] Q. V. Le, "Building high-level features using large scale unsupervised learning," in *2013 IEEE International Conference on Acoustics, Speech and Signal Processing*, 2013, pp. 8595–8598.
[3] S. Gupta, W. Zhang, and F. Wang, "Model Accuracy and Runtime Tradeoff in Distributed Deep Learning: A Systematic Study," in *2016 IEEE 16th International Conference on Data Mining (ICDM)*, 2016, pp. 171–180.
[4] J. Dean *et al.*, "Large scale distributed deep networks," *Proceedings of the 25th International Conference on Neural Information Processing Systems*. Curran Associates Inc., 2012, pp. 1223–1231.
[5] G. W. Burr *et al.*, "Neuromorphic computing using non-volatile memory," *Adv. Phys. X*, vol. 2, no. 1, pp. 89–124, Jan. 2017.
[6] D. E. Rumelhart, G. E. Hinton, and R. J. Williams, "Learning representations by back-propagating errors," *Nature*, vol. 323, no. 6088, pp. 533–536, Oct. 1986.
[7] T. Gokmen and Y. Vlasov, "Acceleration of Deep Neural Network Training with Resistive Cross-Point Devices: Design Considerations," *Front. Neurosci.*, vol. 10, Jul. 2016.
[8] T. Gokmen, O. M. Onen, and W. Haensch, "Training Deep Convolutional Neural Networks with Resistive Cross-Point Devices." [Online]. Available: https://arxiv.org/abs/1705.08014.